\documentclass[prd,nofootinbib,english,superscriptaddress,10pt]{revtex4}
\usepackage{amsfonts,amsmath,hyperref,url, color}
\usepackage{bm, bbm}
\usepackage{graphicx}
\usepackage{mathtools}
\usepackage{nccmath}
\usepackage[font=small,labelfont=bf]{caption}
\usepackage{tabularx}

\usepackage[utf8]{inputenc}
\usepackage[english]{babel}
\usepackage{amsmath,amsfonts,amssymb}
\usepackage{hyperref}
\usepackage{caption,adjustbox}
\usepackage{multirow}
\usepackage{cleveref}
\usepackage{fontawesome}
\usepackage[normalem]{ulem}
\usepackage{comment}
\usepackage{graphicx}
\usepackage{float}
\usepackage{bbding}
\usepackage[usenames,dvipsnames]{xcolor}
\usepackage[arrowdel]{physics}
\usepackage{tensor}
\usepackage{cancel}
\usepackage{tikz}
\usepackage{cleveref}
\usepackage{mathrsfs}
\usepackage{listings} 
\usepackage[title,toc,titletoc]{appendix}
\usepackage{titlesec}
\usepackage{tocloft}
\usetikzlibrary{decorations.pathmorphing, patterns,shapes, decorations.pathreplacing}

	\usepackage{amstext,amsthm,verbatim,cancel,epsfig} 

\usepackage{color}

\newcommand{\be}{\begin{equation}}
\newcommand{\ben}{\begin{equation*}}
\newcommand{\ee}{\end{equation}}
\newcommand{\een}{\end{equation*}}
\newcommand{\ba}{\begin{eqnarray}}
\newcommand{\ea}{\end{eqnarray}}

\newcommand{\bal}{\begin{align}}
\newcommand{\eal}{\end{align}}

\newcommand{\bea}{\begin{eqnarray}}
\newcommand{\eea}{\end{eqnarray}}
\newcommand{\bit}{\begin{itemize}}
\newcommand{\eit}{\end{itemize}}
\newcommand{\la}{\langle}
\newcommand{\ra}{\rangle}

\graphicspath{ {./images} }
\begin{document}

\title{Affine Symmetry and the Group-Theoretic Basis of the Unruh Effect}

\author{Michele Arzano}
\email{michele.arzano@na.infn.it}
\affiliation{Dipartimento di Fisica ``E. Pancini", Universit\`a di Napoli Federico II, I-80125 Napoli, Italy\\}

\affiliation{INFN, Sezione di Napoli, Complesso Universitario di Monte S. Angelo,
Via Cintia Edificio 6, 80126 Napoli, Italy}

\author{Alessandra D'Alise}
\email{alessandra.dalise@na.infn.it}

\affiliation{INFN, Sezione di Napoli, Complesso Universitario di Monte S. Angelo,
Via Cintia Edificio 6, 80126 Napoli, Italy}

\author{Simone del Rosso}
\email{s.delrosso@studenti.unina.it}
\affiliation{Dipartimento di Fisica ``E. Pancini", Universit\`a di Napoli Federico II, I-80125 Napoli, Italy\\}

\author{Domenico Frattulillo}
\email{domenico.frattulillo@na.infn.it}

\affiliation{INFN, Sezione di Napoli, Complesso Universitario di Monte S. Angelo,
Via Cintia Edificio 6, 80126 Napoli, Italy}

\begin{abstract}
A massless scalar field in two spacetime dimensions splits into two independent sectors of left and right-moving modes on the lightcone. At the quantum level, these two sectors carry a representation of the group of affine transformations of the real line, with translations corresponding to transformations generated by light-cone momenta and dilations given by light-cone Rindler momenta formed by a linear combination of generators of boosts and dilations. One-particle states for inertial observers are eigenvectors of translation generators belonging to irreducible representations of the affine group. Rindler one-particle states are related to eigenfunctions of the generator of dilations. We show that simple manipulations connecting these two representations involving the Mellin transform can be used to derive the thermal spectrum of Rindler particles observed by an accelerated observer. Beyond providing a representation-theoretic basis for vacuum thermal effects, our results suggest that analogous phenomena may arise in any quantum system admitting realizations of translation and dilation eigenstates. 
\end{abstract}

\maketitle

\section{Introduction}

The Unruh effect stands as one of the most striking manifestations of the interplay between quantum field theory and the geometry of spacetime. It reveals that the notion of a particle is inherently observer-dependent: a uniformly accelerated observer in Minkowski spacetime perceives the inertial vacuum as a thermal state with temperature proportional to the norm of its acceleration \cite{Unruh}. The  existence of vacuum states that appear “empty’’ only to specific classes of observers is a characteristic feature of quantum fields in non-inertial frames and in general curved spacetimes. This stems from the fact that different observers can adopt different definitions of positive frequency for the field modes, associated to different choices of time-like Killing vectors to generate translations in time, leading to distinct constructions of the one-particle Hilbert space and the associated Fock space \cite{Fulling:1972md, Ashtekar:1975zn}. The celebrated result by Hawking, that black holes radiate thermally with a temperature inversely proportional to their mass \cite{Hawking_1}, and the fact that cosmological horizons also carry an associated temperature \cite{Gibbons:1977mu} are both manifestations of the same underlying phenomenon.  Beyond its significance in horizon thermodynamics, the Unruh temperature of local Rindler horizons plays a central role in Jacobson’s suggestive derivation of Einstein equations from a Clausius relation linking horizon entropy to the energy flux through the horizon \cite{Jacobson:1995ab}. This perspective remains highly relevant in current debates exploring the connection between horizon entropy and space-time dynamics which rely on the same conceptual ingredient: the thermal behavior associated with causal horizons \cite{Jacobson:2015hqa,Svesko:2018qim,Alonso-Serrano:2020pcz,Dorau:2025hmq}.

A wide range of derivations of the Unruh effect exist, each offering a different balance of rigor, simplicity, and physical intuition. Canonical, textbook treatments based on Bogoliubov transformations between Minkowski and Rindler modes clearly expose the origin of the effect due to the exisentce of inequivalent vacuum states but are technically rather involved and offer limited immediate physical intuition.\cite{Unruh,Crisp_Higu}. Detector-based derivations provide an operational interpretation through the excitation spectrum of an Unruh–DeWitt detector, yet depend on idealized pointlike couplings, switching protocols, and subtle regularization schemes \cite{Donoghue:1983, Takagi:1986kn, Louko:2006zv}. Euclidean approaches link thermal behavior to periodic imaginary time or loop contributions, but rest on analytic continuation \cite{Christensen:1978tw,Rajeev:2019bzv} while more algebraic formulations reveal the general KMS structure associated with bifurcate Killing horizons, though at the cost of mathematical abstraction \cite{Sewell:1982zz, Kay:1988mu, Sorce:2024zme}. Other semiclassical, tunneling arguments reproduce the Unruh temperature but typically rely on near-horizon WKB approximations and heuristic identifications of “inside” and “outside” degrees of freedom \cite{Kerner:2006vu}.

In this work we present an additional perspective on the origin of the thermal behavior of quantum fields in the presence of a Rindler horizon. Our approach reveals a representation-theoretic structure underlying the Unruh effect, emerging from the observation that Minkowski and Rindler light-cone translations are intimately related to the group of affine transformations of the real line. From this perspective, Minkowski and Rindler modes emerge as distinct representations which diagobalize different subgroups of the affine group and can be connected through a Mellin transform. We show that inertial one-particle states correspond to eigenstates of translation generators, while Rindler one-particle states are  associated to eigenfunctions of dilation generators within the same affine algebra. The standard Bogoliubov transform between Minkwski and Rindler creation and annihilation operators can be then obtained in terms of a map between these two representations.

In the next section we briefly review a field-theoretic derivation of the Unruh effect in two spacetime dimensions putting emphasis on the decomposition of the field in light-cone coordinates and on the role of the Mellin transform to connect different field decompositions. In Section III we highlight the role of Weyl-Poincarè transformations in the description of Rindler spacetime in two-dimensions and how these translate to affine transformations on the advanced and retarded components of a scalar field on the light-cone. This serves as the main motivation for Section IV in which we discuss the relationship between one-particle states for inertial and Rindler observers and different bases for irreducible representations of the affine group labeled respectively by eigenstates of translation and dilation generators. These two set of states are connected by a Mellin transform which allows to build the group-theoretic analogue of the Bogoliubov transformation. The closing Section V presents a discussion of our results and an outlook to future developments.

\section{Light-cone fields and the Unruh effect in two spacetime dimensions}
\begin{figure}
    \centering
    \includegraphics[width=0.52\linewidth]{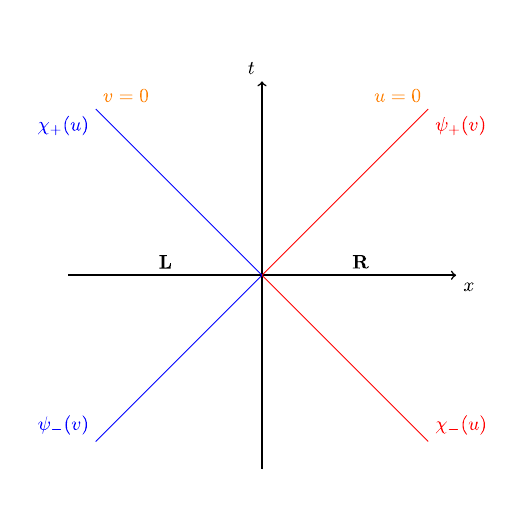}
    \caption{The domains of the four components of $\phi(t,x)$.}
    \label{fig1}
\end{figure}

Let us start from a real massless scalar field $\phi(x,t)$ in 2-dimensional Minkowski spacetime. The Klein-Gordon equation
\be
(\partial_t^2-\partial_x^2)\phi(x,t)=0
\label{wave equation}
\ee
admits solutions which can be written as superpositions of normalized plane waves
\be
\phi(t,x)=\frac{1}{\sqrt{4\pi}}\int^{+\infty}_{-\infty} \frac{dk}{\sqrt{|k|}}[a(k)e^{i(kx-t|k|)}+a^\dagger(k)e^{-i(kx-t|k|)}],
\label{phi iniz}
\ee
where $a(k)$ and $a^\dagger(k)$ are the expansion coefficients which we write in a notation reminiscent of annihilation and creation operators anticipating their roles upon quantization. Introducing light-cone coordinates $u:=x-t$, $v:=x+t$ the general solution \eqref{phi iniz} admits the splitting
\be
\phi(u, v)=\chi(u)+\psi(v)
\label{chi psi}
\ee
where
\be
\psi(v):=\frac{1}{\sqrt{4\pi}}\int^{+\infty}_0\frac{dk}{\sqrt{k}}[a(-k)e^{-ikv}+a^\dagger(-k)e^{ikv}]
\label{psi v}
\ee
and
\be
\chi(u):=\frac{1}{\sqrt{4\pi}}\int^{+\infty}_0\frac{dk}{\sqrt{k}}[a(k)e^{iku}+a^\dagger(k)e^{-iku}]\,.
\label{chi u}
\ee
The components of the decomposition \eqref{chi psi} are independent since eq. \eqref{wave equation} in light-cone coordinates reads $\partial_u\partial_v \phi(u,v)=0$ \cite{libro_wald} whose general solution can be written as the sum of an arbitrary function of 
$u$ and an arbitrary function of $v$. Because these two sectors do not couple in the equation of motion, the theory exhibits a chiral structure with $\chi(u)$ and $\psi(v)$ evolving independently. 
Using the Heaviside step-function, the solutions $\psi(v)$ and $\chi(u)$ can be written as
\be
\psi(v)=:\Theta(v)\psi_+(v)+\Theta(-v)\psi_-(v)
\label{psi hvs}
\ee
and
\be
\chi(u)=:\Theta(u)\chi_+(u)+\Theta(-u)\chi_-(u).
\label{chi hvs}
\ee
Equations (\ref{chi psi}), (\ref{psi hvs}) and (\ref{chi hvs}), 
divide a general solution of the massless Klein-Gordon equation in four contributions. Choosing $u= 0$ or $v= 0$ we have contributions whose domains are \{$u=0,\ v\lessgtr0$\} and \{$v=0,\ u\lessgtr0$\}, as showed in Fig. \ref{fig1}.\\

Uniformly accelerated observers follow hyperbolic trajectories in Minkowski spacetime (see \cite{Arzano:2024ogp} for a pedagogic discussion), which can be described by Rindler coordinates $(\eta,\chi)$, defined in the right (\textbf{R}-) Rindler wedge $|t|<x$ by
\[
t = \frac{e^{\alpha \chi}}{\alpha}\,\sinh(\alpha\eta),\qquad
x = \frac{e^{\alpha \chi}}{\alpha}\,\cosh(\alpha\eta),
\]
where $-\infty< \chi,\, \eta < \infty$ and $\alpha$ is a constant with dimensions of inverse length corresponding to the norm of the four-acceleration of the observer with wordline located at $\chi=0$. Rindler coordinates can also be defined in the left (\textbf{L}-) wedge corresponding to the spacetime region defined by $|t|>x$. As shown in Fig. \ref{fig1}, each wedge is bounded by the domains of the componenets of $\psi(v)$ and $\chi(u)$;
specifically, the \textbf{R}-wedge border is $\{v\geq0\}\ \cup\ \{u\leq0\}$, while for the \textbf{L}-wedge it is $\{v\leq0\}\ \cup\ \{u\geq0\}$. Focusing on $u=const.$ or $v=const.$ subspaces corresponds to considering only one of the independent field components \eqref{psi hvs} and \eqref{chi hvs}. For definiteness, we focus on the \textbf{R}-Rindler wedge and consider the restriction of $\psi(v)$ to positive $v$  
\be
\psi_+(v)=\frac{1}{\sqrt{4\pi}}\int^{+\infty}_0\frac{dk}{\sqrt{k}}[a(-k)e^{-ikv}+a^\dagger(-k)e^{+ikv}], \ \ \ \ v>0.
\label{psi +}
\ee
Introducing the advanced light-cone Rindler coordinate $\xi = \chi+\eta =\frac{1}{\alpha}\log(\alpha v)$ and setting $\alpha=1$, we consider  “Rindler plane waves" plane waves with frequency $\omega$: $e^{i\omega\xi}=v^{i\omega}$. We can rewrite the expression of $\psi_+(v)$ in terms of Rindler plane waves and Rindler frequencies $\omega$ as follows
\be
\psi_+(v)=\frac{1}{\sqrt{4\pi}}\int^{+\infty}_0\frac{d\omega}{\sqrt{\omega}}[v^{-i\omega}B(\omega)+v^{+i\omega}B^\dagger(\omega)].
\label{Rind pla exp}
\ee
where $B(\omega)$ and $B^\dagger(\omega)$ are, again, expansion coefficients which we write as annihilation and creation operators for an uniformly accelerated observer, respectively. In order to obtain an expression of $B(\omega)$ and $B^\dagger(\omega)$ in terms of the annihilation and creation operators  of an inertial observer $a(k)$ and $a^\dagger(k)$ we make use of the Mellin transform of $\psi_+(v)$ (see eg. \cite{Volov}), defined by
\be
\Tilde{\psi}_+(s):=\int^{+\infty}_{0}dv\ v^{s-1}\psi_+(v), \qquad s\in\mathbb{C}.
\label{psi v tilde}
\ee
The inverse trasform of (\ref{psi v tilde})is given by
\be
\psi_+(v)=\frac{v^{-\lambda}}{2\pi}\int^{+\infty}_{-\infty}d\omega\ \Tilde{\psi}_+(s) v^{-i\omega},\qquad s=\lambda+i\omega,
\label{Inv Mell psi}
\ee
The convergence of the integral (\ref{psi v tilde}) requires the parameter $\lambda$ to be restricted in the range $(0,1)$ defining a {\it strip of convergence} in $\mathbb{C}$. We can notice in the expression \eqref{Inv Mell psi} the factor $v^{-i\omega}$ reminiscent of the expansion (\ref{Rind pla exp}) modulo a factor $v^{-\lambda}$. Indeed, the inverse Mellin tranform (\ref{Inv Mell psi}) above is just a different integral representation of  $\psi_+(v)$ and we can obtain the Rindler mode expansion by taking the singular limit $\lambda\to0$. Substituting the expression \eqref{psi +} for $\psi_+(v)$ in \eqref{psi v tilde}, using the integral identity
\be
\int^{+\infty}_0dv\ v^{s-1}e^{\pm ikv}=k^{-s}e^{\pm i\frac{\pi}{2}s}\Gamma(s),
\label{Mellin prop}
\ee
and plugging the resuling expression in \eqref{Inv Mell psi} we can write down 
$B(\omega)$ and $B^\dagger(\omega)$ in terms of the annihilation and creation operators of an inertial observer $a(k)$ and $a^\dagger(k)$
\be
B(\omega,\lambda)=\frac{\sqrt{\omega}}{2\pi}v^{-\lambda}\Gamma(\lambda+i\omega)\int^{+\infty}_{0}\frac{dk}{\sqrt{k}}k^{-i\omega-\lambda}[e^{\frac{\pi}{2}\omega-i\frac{\pi}{2}\lambda}a(-k)
+e^{-\frac{\pi}{2}\omega+i\frac{\pi}{2}\lambda}a^\dagger(-k)]
\label{B omega}
\ee
and
\be
B^\dagger(\omega,\lambda)=\frac{\sqrt{\omega}}{2\pi}v^{-\lambda}\Gamma(\lambda-i\omega)\int^{+\infty}_{0}\frac{dk}{\sqrt{k}}k^{i\omega-\lambda}[e^{-\frac{\pi}{2}\omega-i\frac{\pi}{2}\lambda}a(-k)
+e^{\frac{\pi}{2}\omega+i\frac{\pi}{2}\lambda}a^\dagger(-k)].
\label{B* omega}
\ee
From these expressions one can obtain, using some care in dealing with infrared divergencies, the Unruh thermal distribution in the standard fashion i.e. by evaluating the expectation value of the Rindler number operator $B^\dagger(\omega)B(\omega)$ in the Minkowski vacuum state $\vert0\ra_M$ defined by
\be a(k)\vert0\ra_M=0\,.\ee 
The resulting expression is
\be
\lim_{\lambda\to0}{}_M\la0\vert B^\dagger(\omega,\lambda)B(\omega,\lambda)\vert0\ra_M=\frac{2\pi\delta(0)}{e^{2\pi\omega}-1},
\label{therm dist mink}
\ee
which shows on the right hand side a thermal distribution at temperature $T=\frac{1}{2\pi}$ which, restoring factors of $\alpha$ which we set to unity at the beginning of this section, corresponds with the Unruh temperature $T=\frac{\alpha}{2\pi}$ proportional to the norm of four-acceleration.

The derivation of the Unruh spectrum above may at first appear to be based on a technical shortcut consisting in applying the Mellin transform to quantum fields on the light cone. However, as we show in the following sections, the emergence of the Mellin transform reflects a deeper group-theoretical structure of the Unruh effect, rooted in the role of affine transformations acting on the light cones.

\section{Light-cone evolution and the $ax+b$ group}

Let us have a closer look at how Minkowski and Rindler spaces in two dimensions can be described in terms of groups of transformations. In Minkowski spacetime we have two commuting generators of time and space translations: $P_0=-i\partial_t$ and $P_1=-i\partial_x$ respectively and the group generated
by them acts transitively on Minkowski space i.e. every point of the latter can be reached by transformation of the group of two-dimensional translations generated by $P_0$ and $P_1$. In two-dimensional Rindler spacetime, we also have a transitive action of a group of translations. Such translations consists of boosts and dilations which are generated by $N=-i(x\partial_t+t\partial_x) = -\frac{i}{\alpha}\,\partial_\eta$ and $D=-i(x\partial_x+t\partial_t)= -\frac{i}{\alpha}\,\partial_\chi$ respectively. Taken together, the four generators $\{P_0,P_1,N,D\}$ form the two-dimensional Weyl–Poincaré Lie algebra, characterized by the following non-trivial commutators:
\begin{equation}\label{weyl-poincare}
\begin{aligned}
   & [N,P_0]=i P_1\, ,\qquad [N,P_1]=iP_0\, ,\\
   &[D,P_0]=iP_0\, ,\qquad [D,P_1]=iP_1\, .
    \end{aligned}
\end{equation}
Let us now introduce light-cone generators defined as $P_\pm=P_1\pm P_0$ and $R_\pm=\frac{1}{2}(D\pm N)$. In light-cone coordinates $\{u,v\}$ these generators read 
\begin{equation}\label{pr+-}
    \begin{aligned}
   &P_+=-i\partial_v\, , \qquad R_+=-iv\partial_v\, ,\\
    &P_-=-i\partial_u\, ,\qquad R_-=-iu\partial_u\, .
    \end{aligned}
\end{equation}
and thus $P_{\pm}$ and $R_{\pm}$ generate respectively translations and dilations in advanced and retarded light-cone coordinates. They satisfy the following non-vanishing commutation relations
\begin{equation}
\begin{aligned}
    &[R_+,P_+]=iP_+\, ,\\
    &[R_-,P_-]=iP_-\, ,
    \end{aligned}
\end{equation}
showing a decomposition of the two-dimensional Weyl-Poincaré algebra in two identical sub-algebras \cite{Arzano:2018oby} of the form
\begin{equation}\label{axbalg}
    [R,P]=iP\,. 
\end{equation}
This is the simplest non-abelian Lie algebra and, as we will now discuss, it generates the group of affine transformation of the real line, known also as the {\it “$ax+b$" group}. This decomposition is the algebraic counterpart of the chirality discussed in the previous section: when restricted to null-surfaces defined by $u=const.$ or $v= const.$ Weyl-Poincaré tranformations (this holds true also just for Poincaré transformations) correspond to affine transformations on the real line. Specifically, this is the group $G$ of linear transformations
\be
x\rightarrow ax+b,\ \ \ a\in\mathbb{R}^+,\ b\in\mathbb{R}
\ee
which are combinations of translation and dilation. Denoting by $g:=(a,b)$ the elements of $G$, the multiplication law between $g_1=(a_1,b_1)$ and $g_2=(a_2,b_2)$ is given by
\be
g_1\cdot g_2=(a_1a_2,a_1b_2+b_1).
\label{group molt}
\ee
The $ax+b$-group can be split into two subgroups: $a=1$ defines the subgroup of translations $T(b)=(1,b)$, whose action is transitive on the whole real line  since every $x\in\mathbb{R}$ can be obtained from such transformations by an appropriate choice of $b$. When $b=0$, the group $G$ reduces to the subgroup of dilations $D(a)=(a,0)$; its action is transitive only on the half-line $\mathbb{R}^+$ (or $\mathbb{R}^-$) because the condition $a>0$ implies that a positive (negative) real number can only be mapped into a positive (negative) real number, inducing a partition of the real line.\newline
\begin{figure}
    \centering
    \includegraphics[width=0.6\linewidth]{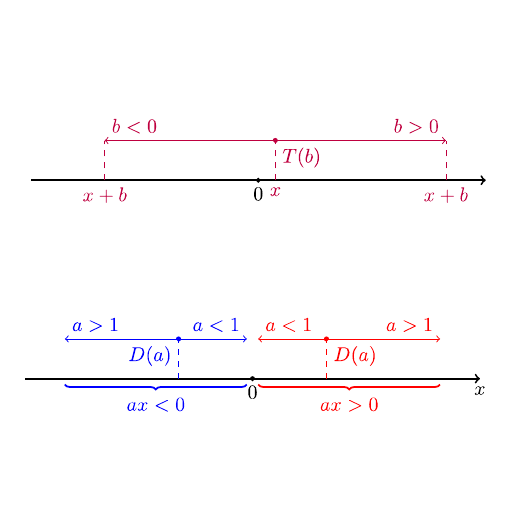}
    \caption{The action of $T(b)$ is transitive on the whole real line; the action of $D(a)$ is transitive only on half-lines.}
    \label{fig2}
\end{figure}

There exist two nontrivial unitary irreducible representations of the affine group \cite{paper_rappr,libr_rapp} that can be realized on the Hilbert spaces of functions $\mathcal{H}^{\pm} = \mathcal{L}^2\left(\mathbb{R}^{\pm}, \frac{dk}{k}\right)$, whose elements can be denoted by kets $\vert k\ra_\pm$ with invariant inner products 
\be
\la\psi\vert\psi'\ra=\int^{+\infty}_0\frac{dk}{k}\psi^*(k)\psi'(k)=\int^{+\infty}_0\frac{dk}{k}\la\psi\vert k\ra_+\ _+\la k\vert\psi'\ra
\ee
and
\be
\la\psi\vert\psi'\ra=\int_{-\infty} ^0\frac{dk}{k}\psi^*(k)\psi'(k)=\int_{-\infty} ^0\frac{dk}{k}\la\psi\vert k\ra_-\ _-\la k\vert\psi'\ra,
\ee
respectively, leading to the orthogonality relation
\be
{}_{\pm}\la k\vert k'\ra_{\pm}= k\, \delta(k-k')\,.
\ee
The actions of translations and dilations on such spaces are given by
\be
T(b)\vert k\ra_\pm=e^{-ib k}\vert k\ra_\pm, \ \ \ \ D(a)\vert k\ra_\pm=\vert a k\ra_\pm\,,
\label{T and D on k}
\ee
showing that dilations move the states along the momentum axis, but they {\it do not mix positive and negative sectors}. Thus, since translations form a normal abelian subgroup, each ket $\vert k\ra_\pm$ defines a character that is carried into another character under dilations, so one obtains full irreducible unitary representations without ever mixing the positive and negative momentum sectors. This construction is a special case of Mackey's theory of induced representations \cite{raczka1986theory}.
We can write the actions \eqref{T and D on k} in terms of generators, which we denote with $P$ and $R$, eqs. (\ref{T and D on k}) are given by
\be
T(\beta)\vert k\ra_\pm=e^{-i\beta P}\vert k\ra_\pm=e^{-i\beta k}\vert k\ra_\pm
\label{T and P}
\ee
and
\be
D(\alpha)\vert k\ra_\pm=e^{-i \alpha R}\vert k\ra_\pm=\vert e^\alpha k\ra_\pm.
\label{D and R}
\ee
Differentiating with respect to the parameters $\alpha$ and $\beta$, one obtains the action of the generators on $\vert k\ra_\pm$
\be
P\vert k\ra_\pm=k\vert k\ra_\pm 
\label{P on k}
\ee
and
\be
R\vert k\ra_\pm=-ik\frac{d}{dk}\vert k\ra_\pm,
\label{R on k}
\ee
hence the $\vert k\ra_\pm$ states are eigenstates of the P-operator of eigenvalue $k\in\mathbb{R}^\pm$. Using the actions (\ref{P on k}) and (\ref{R on k}) one can obatin the commutator \eqref{axbalg}, as expected.

\section{Inequivalent Minkowski and Rindler modes in one-dimensional affine group representations}

We will now consider the representations of the affine group above as the Hilbert spaces of a quantum system. Such system can be identified with a 0+1-dimensional quantum field theory or as the states of a one-dimensional quantum system with affine symmetry \cite{proc_2,proc_1}. Since the unitary transformations (\ref{T and D on k}) can be seen as the light-cone counterpart of Poincaré transformations, the states $\vert k\ra_+$, eigenstates of the generator of translations, $P$ can be seen as one-particle states which we dub “P-particle states". In terms of such P-eigenstates the $0+1$-dimensional affine field $\psi(t)=\la t\vert\psi\ra$ can be expanded as \cite{MosesQuesada1972ScaleEuclidean}
\be
\psi(t)=\la t\vert\psi\ra=\frac{1}{\sqrt{2\pi}}\int^{+\infty}_0\frac{dk}{k}e^{ikt}\ _+\la k\vert\psi\ra+\frac{1}{\sqrt{2\pi}}\int^0_{-\infty}\frac{dk}{|k|}e^{ikt}\ _-\la k\vert\psi\ra
\label{psi t k}
\ee
defining a reducible representation of the affine group on the real line. Indeed, irreducible representations on the whole line $t\in\mathbb{R}$ are defined considering only positive or negative frequency $k$. Moreover, we observe that the bra-kets
\be
\la t\vert k\ra_+=\frac{1}{\sqrt{2\pi}}e^{ikt}
\ee
can be interpreted as the counterparts of plane waves representing one particle states in quantum field theory. In fact we can rewrite the expansion \eqref{psi t k} as
\be
\psi(t)=\frac{1}{\sqrt{2\pi}}\int^{+\infty}_0\frac{dk}{k}(e^{ikt}\ _+\la k\vert\psi\ra+ e^{-ikt}\ _-\la-k\vert\psi\ra).
\ee
 Requiring the reality condition $_-\la-k\vert\psi\ra=(_+\la k\vert\psi\ra^*)$ yields to
\be
\psi(t)=\frac{1}{\sqrt{2\pi}}\int^{+\infty}_0\frac{dk}{k}[e^{ikt}\ _+\la k\vert\psi\ra+e^{-ikt}\ (_+\la k\vert\psi\ra)^*].
\label{psi t 2}
\ee
Looking at the expansion coefficients of the field
\be
a(k):={}_+\la k\vert\psi\ra,
\label{a in real line}
\ee
we have that
\be
\la k\vert P\vert\psi\ra=k\la k\vert\psi\ra=ka(k)\Rightarrow T(\beta)a(k)=e^{-i\beta P}a(k)=e^{i\beta k}a(k).
\label{T on a}
\ee
and, consequently, the action of the translation operator $T(\beta)$ on (\ref{psi t 2}) reads as
\be
T(\beta)\psi(t)=\psi(t+\beta)
\label{T on psi}
\ee
and similarly, the action of $D(\alpha)$ on $\psi(t)$ is given by
\be
D(\alpha)\psi(t)=\psi(e^\alpha t)\,.
\label{D on psi}
\ee
From (\ref{T on psi}) and (\ref{D on psi}) it is clear that time evolution, generated by $P$, covers the whole real line while transformations generated by $R$ cover only the positive or the negative half-lines.\\ 

The expression \eqref{psi t 2} for the affine quantum field, is formally identical to the advanced light-cone component (\ref{psi v}) of the real field in Minkowski spacetime. Since the same identification holds for the retarded light-cone component \eqref{chi u} we see that we can fully describe a free real quantum field in two-dimensional Minkwoski spacetime in terms of two copies of an affine field constructed from representations of the $ax+b$ group. We thus expect that the {\it existence of inequivalent notions of particle excitations} associated to inertial and uniformly accelerated observers in Minkowski space-time can be captured already within the simple one-dimensional framework of affine group representations.\\ 

To see that this is indeed the case, let us introduce a new coordinate $\xi$ “adapted" to dilations i.e. such that $D(\lambda)$ acts on it as a ordinary translation. Considering (\ref{D on psi}), it is clear that such change of variable is given by
\be
t=e^\xi\,
\label{txi}
\ee 
from which
\be
D(\alpha)\, \psi(\xi)=\psi(\xi+\alpha).
\label{D on psi xi}
\ee
The action of standard translations on $\psi(\xi)$ becomes less trivial:
\be
T(\beta)\, \psi(\xi)=\psi(\log(e^\xi+\beta)).
\label{T on psi xi}
\ee
From (\ref{D on psi xi}) and (\ref{T on psi xi}) one can derive the expressions of P and R in terms of $\xi$:
\be
R=-i\frac{d}{d\xi},\ \ \ \ P=-ie^{-\xi}\frac{d}{d\xi},
\label{R and P xi}
\ee
which are a representation of the algebra (\ref{axbalg}) over the positive half-line $t\in\mathbb{R}^+$ ($\xi\in\mathbb{R}$).\\ 

The expansion \eqref{psi t k} of the affine field in terms of modes associated to the $|k\ra_{\pm}$ bases in the irreducible representations of the $ax+b$ group diagonalize the subgroup of translations. Under the partition of the configuration $t$-line induced by the change of variable \eqref{txi}, it is natural to look at the expansion of the affine field in terms of modes which are plane waves oscillating with R-frequency $\omega$ 
\be
\la \xi| \omega\ra_{\xi+} = \frac{1}{\sqrt{2\pi}}e^{i\omega\xi}=\frac{1}{\sqrt{2\pi}}t^{i\omega}
\qquad t\in\mathbb{R}^+,
\label{omega R}
\ee
where the states $| \omega\ra_{\xi+}$ provide a spectral resolution of $R$ on the positive half of configuration space $t\in\mathbb{R}^+$. One can easily derive from \eqref{R and P xi} the action of $R$ and $P$ on the $| \omega\ra_{\xi+}$ states
\be
R\,\vert\omega\ra_{\xi+}=\omega \vert\omega\ra_{\xi+}, \qquad P\,\vert\omega\ra_{\xi+}=\omega\vert\omega+i\ra_{\xi+}\,,
\ee
showing that, while $R$ is diagonal on such states, $P$ is not diagonal and its action produces a complex shift in the $\xi$ frequency.

Since the plane waves $e^{i\omega\xi}$ provide a complete set of functions for $\xi\in\mathbb{R}$, i.e., $t\in\mathbb{R}^+$, the functions $t^{i\omega}$ form a complete set on the positive half-line. Analogously, on the negative half-line $\xi$, the plane waves $e^{i\omega\xi}=(-t)^{i\omega}$ form a complete set. In terms of these new plane waves the affine field $\psi(t)$ can be expanded as
\be
\psi(t)=\Theta(t)\psi_+(t)+\Theta(-t)\psi_-(t)=\frac{1}{\sqrt{2\pi}}\int^{\infty}_0 d\omega(t^{i\omega}c_+(\omega)+t^{-i\omega}c_+^*(\omega))
+\frac{1}{\sqrt{2\pi}}\int^{\infty}_0d \omega((-t)^{-i\omega}c_-(\omega)+(-t)^{+i\omega}c_-^*(\omega))
\label{psi t pos e neg}
\ee
where $c_\pm(\omega) = \la \psi|\omega\ra_{\xi\pm}$ are the expansion coefficients in terms of the $R$-plane waves. Notice how the partition induced by the introduction of the $\xi$ configuration variable breaks translational symmetry and thus the modes in the expansion \eqref{psi t pos e neg} {\it do not} carry a unitary representation of the $ax+b$ group, albeit being eigenfunctions of the differential operator $R=-i\frac{d}{d\xi}$.\\

Eigenstates of the generators of dilations $R$ can be alternatively introduced through a Mellin transform of the eigenstates of $P$ with a pure imaginary variable $i\omega$
\be
\vert\omega\ra=\frac{1}{\sqrt{2\pi}}\int^{+\infty}_0\frac{dk}{k}k^{-i\omega}\vert k\ra_+\,,
\label{omega lambda 0}
\ee
with
\be
\la\omega\vert k\ra_+=\frac{1}{\sqrt{2\pi}}\, k^{i\omega}.
\label{omegak}
\ee
The Mellin transform \eqref{omega lambda 0} is a unitary change of basis to the kets $|\omega\ra$ on the Hilbert space carrying an irreducible representation of the $ax+b$ group, with the label $\omega$ now running over the entire real line representing functions belonging to the Hilbert space $\mathcal{H}= \mathcal{L}^2(\mathbb{R},\, d\omega)$. Using eqs. (\ref{P on k}) and (\ref{R on k}), the actions of $P$ and $R$ on \eqref{omega lambda 0} read
\be
R\, \vert\omega\ra=\omega\vert\omega\ra, \qquad P\, \vert\omega\ra=\vert\omega\textcolor{black}{+}i\ra\,, 
\label{R e P omega}
\ee
showing thet the new basis is diagonal under the action of the generator of dilations $R$. One can easily check that, while the action of finite dilations on (\ref{omega lambda 0}) is diagonal
\be
D(\alpha)\, \vert\omega\ra=e^{i \alpha\omega}\vert\omega\ra,
\ee
for translations we have \cite{libr_rapp}
\be
T(\beta)\, \vert\omega\ra=\frac{1}{\sqrt{2\pi}}\int^{+\infty}_0\frac{dk}{k}\ k^{\textcolor{black}{-}i\omega}e^{-i\beta k}\vert k\ra_+
\ee
which, using the inverse Mellin transform 
\begin{equation}
|k\rangle_+=\frac{1}{\sqrt{2\pi}}\int_{-\infty}^{\infty} d\omega'\;k^{\textcolor{black}{+}i\omega'}\,|\omega'\rangle ,
\end{equation}
becomes
\begin{equation}
T(\beta)|\omega\rangle
=\int_{-\infty}^{\infty} d\omega'\;
\left[
\frac{1}{2\pi}\int_0^\infty\frac{dk}{k}\;
k^{\,\textcolor{black}{-}i(\omega-\omega')}\,e^{-i\beta k}
\right]\,|\omega'\rangle .
\end{equation}
Thus the kernel of the transformation is
\begin{equation}
K(\omega,\omega';\beta)
=\frac{1}{2\pi}
\int_0^\infty\frac{dk}{k}\;
k^{\textcolor{black}{-}\,i(\omega-\omega')}\,e^{-i\beta k}.
\end{equation}
Evaluating the integral with
\begin{equation}
\int_0^\infty x^{s-1}e^{-a x}dx
= a^{-s}\Gamma(s),
\qquad (\Re a>0),
\end{equation}
and analytically continuing to $a=i\beta$, $s=\textcolor{black}{-}i(\omega-\omega')$, we find
\begin{equation}
K(\omega,\omega';\beta)
=\frac{1}{2\pi}\,
(i\beta)^{\textcolor{black}{+}\,i(\omega-\omega')}\,
\Gamma\!\big(\textcolor{black}{-}i(\omega-\omega')\big).
\end{equation}
Therefore the action of translations on the Mellin basis is
\begin{equation}
T(\beta)\,|\omega\rangle
= \int_{-\infty}^{\infty} d\omega'\;
K(\omega,\omega';\beta)\;
|\omega'\rangle 
\end{equation}
showing that translations mix eigenstates of dilations with positive and negative frequencies.\\

In order to find the relationship between the expansion coefficients $c_\pm(\omega) = {}_{\xi\pm}\la \omega| \psi\ra$ and the coefficients of the standard plane wave expansion $a(k):={}_+\la k\vert\psi\ra$, we consider the expression for the affine field $\psi(t)$ in eq. \eqref{psi t 2} and rewrite it as
\be
\psi(t)=\frac{1}{2\pi}\int^{+\infty}_{-\infty}\textcolor{black}{d\omega}\int^{+\infty}_{0}\frac{dk}{k}(e^{ikt}k^{\textcolor{black}{-}i\omega} \la\omega\vert\psi\ra+e^{-ikt}k^{\textcolor{black}{+}i\omega}\la\omega\vert\psi\ra^*),
\label{psi non regular}
\ee
where we have used eq.\eqref{omegak} and the completeness relation
\be
1=\int^{+\infty}_{-\infty}d\omega\vert\omega\ra\la\omega\vert.
\ee
We notice that the integral in eq.\eqref{psi non regular} is not defined for $\omega\in\mathbb{R}$, hence we regularize it by promoting $\omega$ to a complex variable considering $\omega\rightarrow\omega\textcolor{black}{+}i\lambda$ with $\lambda\in(0,1)$.
\be
\psi(t)=\frac{1}{2\pi}\int^{+\infty}_0\frac{dk}{k}\int^{+\infty}_{-\infty}d\omega(e^{ikt}k^{\textcolor{black}{-}\lambda\textcolor{black}{-}i\omega}\ \la\omega\vert\psi\ra+e^{-ikt}k^{\textcolor{black}{-}\lambda\textcolor{black}{+}i\omega}\ \la\omega\vert\psi\ra^*).
\ee
Integrating with respect to $k$ we obtain
\begin{equation}
\begin{aligned}
    \psi(t)&=\Theta(t)\frac{1}{2\pi}\int^{+\infty}_{-\infty}d\omega\biggl[\Gamma(\textcolor{black}{-}\lambda\textcolor{black}{-}i\omega)t^{\textcolor{black}{+}\lambda\textcolor{black}{+}i\omega}e^{i\frac{\pi}{2}(\textcolor{black}{-}\lambda\textcolor{black}{-}i\omega)}\ \la\omega\vert\psi\ra +\Gamma(\textcolor{black}{-}\lambda\textcolor{black}{+}i\omega)t^{\textcolor{black}{+}\lambda\textcolor{black}{-}i\omega}e^{-i\frac{\pi}{2}(\textcolor{black}{-}\lambda\textcolor{black}{+}i\omega)}\ \la\omega\vert\psi\ra^*\biggr]\\
    & +\Theta(-t)\frac{1}{2\pi}\int^{+\infty}_{-\infty}d\omega\biggl[\Gamma(\textcolor{black}{-}\lambda\textcolor{black}{-}i\omega)(-t)^{\textcolor{black}{+}\lambda\textcolor{black}{+}i\omega}e^{-i\frac{\pi}{2}(\textcolor{black}{-}\lambda\textcolor{black}{-}i\omega)}\ \la\omega\vert\psi\ra+\Gamma(\textcolor{black}{-}\lambda\textcolor{black}{+}i\omega)(-t)^{\textcolor{black}{+}\lambda\textcolor{black}{-}i\omega}e^{+i\frac{\pi}{2}(\textcolor{black}{-}\lambda\textcolor{black}{+}i\omega)}\ \la\omega\vert\psi\ra^*\biggr].
\end{aligned}
\end{equation}
Dividing the domains of integration into $(-\infty,0)$ and $(0,+\infty)$ and changing the sign of the variable $\omega$ in the integrals on the negative half-line, one can write
\begin{equation}
\begin{aligned}
\psi(t)&=\Theta(t)\Biggl[\int^{+\infty}_0\frac{d\omega}{2\pi}\ t^{\textcolor{black}{+}\lambda}\Gamma(\textcolor{black}{-}\lambda\textcolor{black}{-}i\omega)t^{\textcolor{black}{+}i\omega}[e^{\textcolor{black}{-}i\frac{\pi}{2}\lambda\textcolor{black}{+}\frac{\pi}{2}\omega}\ b_+(\omega)+e^{\textcolor{black}{+}i\frac{\pi}{2}\lambda\textcolor{black}{-}\frac{\pi}{2}\omega}\ b_-^*(\omega)]\\ &+\int^{+\infty}_0\frac{d\omega}{2\pi}\ t^{\textcolor{black}{+}\lambda}\Gamma(\textcolor{black}{-}\lambda\textcolor{black}{+}i\omega)t^{\textcolor{black}{-}i\omega}[e^{\textcolor{black}{+}i\frac{\pi}{2}\lambda\textcolor{black}{+}\frac{\pi}{2}\omega}\ b_+^*(\omega)+e^{+i\frac{\textcolor{black}{-}\pi}{2}\lambda\textcolor{black}{-}\frac{\pi}{2}\omega}\ b_-(\omega)]\Biggr]\\
& +\Theta(-t)\Biggl[\int^{+\infty}_0\frac{d\omega}{2\pi}\ (-t)^{\textcolor{black}{+}\lambda}\Gamma(\textcolor{black}{-}\lambda\textcolor{black}{-}i\omega)(-t)^{\textcolor{black}{+}i\omega} [e^{\textcolor{black}{+}i\frac{\pi}{2}\lambda\textcolor{black}{-}\frac{\pi}{2}\omega}\ b_+(\omega)+e^{\textcolor{black}{-}i\frac{\pi}{2}\lambda\textcolor{black}{+}\frac{\pi}{2}\omega}\ b_-^*(\omega)]\\&+\int^{+\infty}_0\frac{d\omega}{2\pi}\ (-t)^{\textcolor{black}{+}\lambda}\Gamma(\textcolor{black}{-}\lambda\textcolor{black}{+}i\omega)(-t)^{\textcolor{black}{-}i\omega}[e^{\textcolor{black}{-}i\frac{\pi}{2}\lambda\textcolor{black}{-}\frac{\pi}{2}\omega}\ b_+^*(\omega)+e^{\textcolor{black}{+}i\frac{\pi}{2}\lambda\textcolor{black}{+}\frac{\pi}{2}\omega}\ b_-(\omega)]\Biggr],
\label{psi from b lambda}
\end{aligned}
\end{equation}
where we used the bra version of \eqref{omega lambda 0} to define 
\be
b_{\pm}(\omega,\lambda)=\frac{1}{\sqrt{2\pi}}\int^{+\infty}_0\frac{dk}{k}k^{\textcolor{black}{\pm} i(\omega-i\lambda)}a(k)\,.
\label{b regular}
\ee

As for field quantization in advanced and retarded light-cone coordinates we can construct a Fock space for the affine field by promoting the expansion coefficients to creation and annihilation operators. The P-plane wave expansion coefficient then defines a vacuum state $|0\ra_P$ such that 
\begin{equation}
    a(k)\, |0\ra_P = 0\qquad \forall\,k\,
\end{equation}
and thus also
\begin{equation}
    b_{\pm}(\omega,\lambda)\, |0\ra_P = 0\qquad \forall\,\omega\,
\end{equation}
Focussing, for definiteness, on the positive half line component of (\ref{psi t pos e neg}) we have then 
\be
\psi_+(t)=\frac{1}{\sqrt{2\pi}}\int^{+\infty}_0 d\omega\,(t^{i\omega}c_+(\omega)+t^{-i\omega}c_+^\dagger(\omega))
\label{psi t plus}
\ee
and comparing Eq. (\ref{psi from b lambda}) with Eq. (\ref{psi t plus}) and taking the limit $\lambda\to0^+$, we obtain
\be
c_+(\omega)=\frac{1}{\sqrt{2\pi}}\Gamma(\textcolor{black}{+}i\omega)[e^{-\frac{\pi}{2}\omega}\ b_{\textcolor{black}{-}}^\dagger(\omega)+e^{+\frac{\pi}{2}\omega}\ b_{\textcolor{black}{+}}(\omega)]
\label{c omega}
\ee
which is the real line counterpart of the operator $B_+(\omega)$ in (\ref{B omega}). Using familiar manipulations, we can show that the expectation value of the number operator for the Mellin frequency particles $c_+(\omega)c^{\dagger}_+(\omega)$ in the P-vacuum is given by
\begin{figure}
    \centering
    \includegraphics[width=0.5\linewidth]{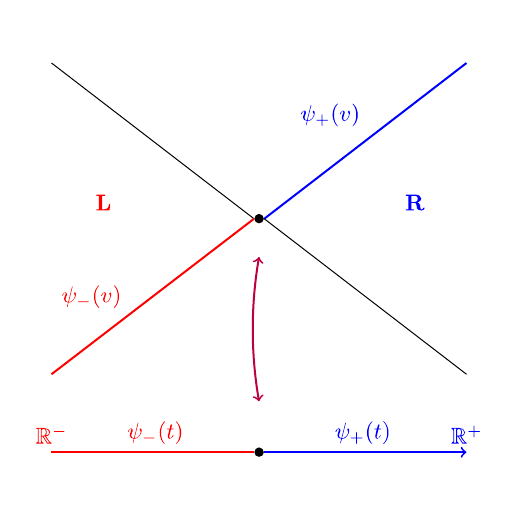}
    \caption{The expression (\ref{psi t pos e neg}) is analogous to the expansion of $\psi(v)$ (or $\chi(u)$), component of the 2-dimensional massless scalar field in Minkowski spacetime, into $\psi_+(v)$ and $\psi_-(v)$ (\ref{psi hvs}) (or $\chi_-(u)$ and $\chi_+(u)$ (\ref{chi hvs}), respectively). In this sense, the division of Rindler spacetime into two wedges corresponds to the partition of the real line induced by $D(\lambda)$.}
    \label{fig3}
\end{figure}

\be
\lim_{\lambda\to0}\la0\vert c_+^\dagger(\omega,\lambda)c_+(\omega)\vert0\ra=\la0\vert c_+^\dagger(\omega)c_+(\omega)\vert0\ra=\frac{\delta(0)}{e^{2\pi\omega}-1},
\label{therm dist real line}
\ee
namely a thermal distributio at temperature $T=\frac{1}{2\pi}$.\\

Let us now wrap up our analysis from a representation-theoretic point of view. The eigenstates $|\omega\rangle$ of the dilation generator $R$ arise as a Mellin transform of the translation eigenstates $|k\rangle_+$ and belong to the Hilbert space 
$\mathcal H = \mathcal{L}^2(\mathbb R, d\omega)$ carrying a unitary irreducible representation of the affine group. Explicitly,
\begin{equation}
|\omega\rangle = \frac{1}{\sqrt{2\pi}}\int_0^\infty \frac{dk}{k}\, k^{\textcolor{black}{-}i\omega}\,|k\rangle_+,
\qquad
\langle \omega|\omega'\rangle=\delta(\omega-\omega')\,.
\end{equation}
In this basis, translations mix different $\omega$ modes through a unitary kernel,
\begin{equation}
T(\beta)|\omega\rangle
=\int_{-\infty}^{\infty}d\omega'\,
K(\omega,\omega';\beta)\,|\omega'\rangle ,
\end{equation}
so that frequency mixing occurs but unitarity is preserved, as required within a single irreducible representation. On the other hand, the $\xi$--plane waves
\begin{equation}
\langle \xi|\omega\rangle_{\xi+}
=\frac{1}{\sqrt{2\pi}}e^{i\omega\xi},
\qquad \xi\in\mathbb R,
\end{equation}
diagonalize the differential operator $R=-i\partial_\xi$ acting on functions of $\xi$, and form an orthonormal basis of $\mathcal{L}^2(\mathbb R,d\xi)$:
\be
{}_{\xi+} \la \omega | \omega' \ra_{\xi +} = \delta (\omega - \omega')\,.
\label{xinner}
\ee
These states define a representation of the dilation subgroup only. Indeed, while
\begin{equation}
R\,|\omega\rangle_{\xi+}=\omega|\omega\rangle_{\xi+} ,
\end{equation}
the action of translations generated by $P=-ie^{-\xi}\partial_\xi$ is not diagonal and does not preserve the $\mathcal{L}^2(\mathbb R,d\xi)$ inner product.

The connection between the two $\omega$-labelled states is obtained by projecting the Mellin eigenstates onto the $\xi$-representation:
\begin{equation}
{}_+\langle \xi|\omega\rangle
=\frac{1}{\sqrt{2\pi}}\int_0^\infty\frac{dk}{k}\,k^{\textcolor{black}{-}i\omega}e^{ik e^\xi}
=\frac{1}{\sqrt{2\pi}}\,e^{i\omega\xi}\,
e^{\textcolor{black}{+}\frac{\pi\omega}{2}}\Gamma(\textcolor{black}{-}i\omega),
\end{equation}
up to a phase convention. 
When the affine-invariant inner product is expressed in the basis of $\xi$-plane waves, one finds
\be
\langle \omega|\omega'\rangle =
|\Gamma(\textcolor{black}{-}i\omega)|^2\,e^{\pi\omega}\,\delta(\omega-\omega').
\ee
The additional $\omega$-dependent factor on the right hand side renders this expression different from the plain $\delta$-function defining the natural inner product \eqref{xinner} on $L^2(\mathbb R,d\xi)$, with respect to which the plane waves $|\omega\rangle_\xi$ are orthonormal. Consequently, the map
\be
|\omega\rangle \;\longmapsto\; |\omega\rangle_{\xi+}
\ee
does not preserve norms and does not intertwine the unitary action of the affine group: it is therefore {\it not unitary}. In summary, only the Mellin states $|\omega\rangle$ belong to a unitary irreducible representation of the affine group. The $\xi$--plane waves associated to the states $|\omega\rangle_{\xi+}$, while natural as modes diagonalizing $R$ on the positive half of configuration space, do not carry a unitary representation of the full symmetry group. This representation-theoretic mismatch is the origin of the inequivalence between inertial (Mellin) and Rindler notions of particles.

\section{Summary}
We studied an unexplored connection between the representation theory of the $ax+b$ group and the light-cone description of the Unruh effect. We illustrated how the retarded and advanced components of a massless Klein-Gordon field can be understood in terms of representations of the affine group on the real line. This can be seen as the one-dimensional counterpart of the familiar description of a quantum field in terms of representations of the Poincaré group and indeed the $ax+b$ group is the simplest Lie group with a semi-direct structure whose representations can be built starting from the one-dimensional representations of the normal sub-group of translations.

Our analysis has two main points of significance. First, it provides a deeper understanding of the group-theoretic structures underlying the Unruh effect and, more generally, the thermodynamic behaviour associated with causal horizons. By casting the transition between inertial and Rindler descriptions as a change of basis between inequivalent realizations of affine-group representations, the appearance of a thermal spectrum becomes a direct consequence of the underlying symmetry structure. Second, the construction we presented can be relevant to non-relativistic quantum systems whose states can be described using representations of the affine group. In particular, systems in which one can experimentally access eigenstates of either the translation generator or the dilation generator provide a natural setting in which the mathematics underlying the Unruh effect reappears. In such contexts there arises the possibility of probing thermal features analogous to those found in semiclassical gravity and quantum field theory in curved spacetime. Determining whether these systems can display experimental signatures of Unruh-like thermality is a fascinating direction for future work. This possibility, together with the broader implications of the analysis we presented, is left to upcoming studies.

\section*{Acknowledgements}
MA and DF acknowledge support from the INFN Iniziativa Specifica QUAGRAP. 
    
\bibliography{bibliography.bib}
\bibliographystyle{utphys}

\end{document}